\documentclass[12pt]{article}
\usepackage{amsmath,amssymb,bm,graphicx}
\usepackage{color} 
\usepackage{hyperref}

\newcommand{\delslash}{\not \! \partial}

\begin{document}

\begin{flushright}
\end{flushright}

\begin{center}
{\Large{\bf Neutrinoless double $\beta$ decay and leptogenesis in seesaw model}}
\end{center}
\vskip .5 truecm
\begin{center}
{\bf { Kazuo Fujikawa 
}}
\end{center}
\begin{center}
\vspace*{0.4cm} 
{\it {Center for Interdisciplinary Theoretical and
Mathematical Sciences,\\
RIKEN, Wako 351-0198, Japan}
}
\end{center}
\makeatletter
\makeatother


\begin{abstract}
The fermion 
$\nu_{R}+C\overline{\nu_{R}}^{T}$ in the  Type I  seesaw model, which is commonly identified with a Majorana fermion,  is not a  Majorana fermion and thus no neutrinoless double $\beta$ decay (in the case of $\nu_{L}-C\overline{\nu_{L}}^{T}$) due to the theorem on the absence of a Majorana-Weyl fermion in the representation of   the d=4 Lorentz group. The  leptogenesis is 
realized naturally together with the neutrinoless double $\beta$ decay by first defining C symmetric Majorana
fermions  in the seesaw model using an analogue of the Bogoliubov transformation. 

\end{abstract}

 \section{ Introduction}
 
Recently, it has been shown that the original definition 
given by
E. Majorana \cite{Majorana} is the only consistent
definition of Majorana fermions in d=4 and that the neutrinoless
double $\beta$ decay is described only by those fermions; this
analysis is based  on the theorem on the absence of
Majorana-Weyl fermions in the representation of the d=4 proper Lorentz group \cite{Fujikawa}, unlike d=2 or 10, for example.  

It is important to recall the consistent definitions of
C, P and CP in field theory.
The representation with $\gamma_{5}$ diagonal is convenient for this 
purpose; the $\gamma$ matrices are then
defined by \cite{Schechter}, 
\begin{eqnarray}\label{metric conventions}
\vec{\gamma}=\left(\begin{array}{cc}
            0&-i\vec{\sigma}\\
            i\vec{\sigma}&0
            \end{array}\right),
\gamma_{4}=\left(\begin{array}{cc}
            0&1\\
            1&0
            \end{array}\right),  
\gamma_{5}=\left(\begin{array}{cc}
            1&0\\
            0&-1
            \end{array}\right),  
C=\left(\begin{array}{cc}
            -\sigma_{2}&0\\
            0&\sigma_{2}
            \end{array}\right)
\end{eqnarray}  
and 
the 4-component Dirac-type field is parameterized without assuming 
a specific form of the Dirac equation by 
\begin{eqnarray}\label{4-component Dirac}
&&\psi(x) =\left(\begin{array}{c}
            \chi\\
            \sigma_{2}\phi^{\star}
            \end{array}\right) , \ \  
\psi^{C}(x) =C\overline{\psi}^{T}(x)=\left(\begin{array}{c}
            \phi\\
            \sigma_{2}\chi^{\star}
            \end{array}\right).     
 \end{eqnarray}  
 The fields with the parameterization \eqref{4-component Dirac} 
without specific Dirac equations are called Dirac-type fields in the 
present paper.
 The above C transformation implies in terms of two-component 
spinors
 \begin{eqnarray}\label{C of two-spinor}
  \hat{C}:\ \chi \rightarrow \phi,\ \ \phi \rightarrow \chi,
  \end{eqnarray}
 namely, in the chiral notation 
 \begin{eqnarray}
\hat{C}:\   \psi_{R}=\left(\begin{array}{c}
            \chi\\
            0
            \end{array}\right) \rightarrow                  
 C\overline{\psi_{L}}^{T}=\left(\begin{array}{c}
            \phi\\
            0
            \end{array}\right), \
 \psi_{L}=\left(\begin{array}{c}
            0\\
            \sigma_{2}\phi^{\star}
            \end{array}\right) \rightarrow                  
 C\overline{\psi_{R}}^{T}=\left(\begin{array}{c}
            0\\
            \sigma_{2}\chi^{\star}
            \end{array}\right) .
\end{eqnarray}
We thus  abstract the C and P transformation laws of chiral fermions as        
\begin{eqnarray}\label{C and P}
\hat{C}&:& \ \nu_{L}(x)\rightarrow \nu^{C}_{L}(x)=C\overline{\nu_{R}}^{T}(x), \ \ \nu_{R}(x)\rightarrow \nu^{C}_{R}(x)= C\overline{\nu_{L}}^{T}(x),\nonumber\\
\hat{P}&:& \ \nu_{L}(x)\rightarrow\nu^{P}_{L}(x)= i\gamma_{4}\nu_{R}(t,-\vec{x}), \ \ \nu_{R}(x)\rightarrow \nu^{P}_{R}(x) =i\gamma_{4}\nu_{L}(t,-\vec{x}).
\end{eqnarray}
where the parity is defined by the chiral projections of $\psi(x)^{P}=i\gamma_{4}\psi(t,-\vec{x})$, and the combined CP is defined by 
\begin{eqnarray}\label{CP}
\hat{C}\hat{P}:\  \nu_{L}(x)\rightarrow\nu^{CP}_{L}(x)= i\gamma_{4}C\overline{\nu_{L}}^{T}(t,-\vec{x}), \ \ \nu_{R}(x)\rightarrow \nu^{CP}_{R}(x)= i\gamma_{4}C\overline{\nu_{R}}^{T}(t,-\vec{x}).
\end{eqnarray}
The extra $i$ for the parity (mirror symmetry) which is characteristic 
to the Majorana fermion together with a modification of T-reversal, as was noted by Majorana himself  
(footnote 2 in his paper \cite{Majorana}), is incorporated. 
This extra $i$ is applied to all the  fermions but it does not 
influence the ordinary fermion number conserving terms.  
 The original Majorana
fermions  are defined by  \cite{Majorana}
\begin{eqnarray}\label{Majorana}
\psi_{M_{+}}(x)= \frac{1}{\sqrt{2}}[\psi(x)+ C\overline{\psi(x)}^{T}], \ \ 
\psi_{M_{-}}(x)&=&\frac{1}{\sqrt{2}}[\psi(x)- C\overline{\psi(x)}^{T}]
\end{eqnarray}
in terms of a Dirac-type fermion $\psi(x)$; it is also possible to divide $\psi_{M_{-}}$ by the factor $i$. One may recall that the 
CPT theorem \cite{CPT} was formulated by analyzing the theories with Dirac 
fermions  before the discovery of parity violation.  The definition of 
$\psi_{M_{+}}$ is consistent with that  of Pauli \cite{Pauli}, who did not treat chiral fermions explicitly.

The above chirality preserving representation of the charge 
conjugation $\hat{C}$ with the doublet $(\nu_{R},\nu_{L})$ 
(a Dirac-type fermion) is consistent in field theory and used for all the fermions in the 
Standard Model (SM); for example, the anti-neutrino is specified by CP not 
by C when the neutrino $\nu_{L}$ is understood as a Weyl fermion
since $\nu_{R}$ 
is then absent (i.e., intrinsic parity violation).
A crucial implication of these 
definitions is that 
the commonly used "Majorana" particle in the Type I seesaw model (an 
extension of the SM)
\begin{eqnarray}\label{pseudoMajorana1} 
\psi_{+}(x)&=& \nu_{R}+ C\overline{\nu_{R}}^{T}
\end{eqnarray}
is {\em not} a Majorana fermion in the sense of $\hat{C}$ in 
\eqref{C and P}. The fermion 
\eqref{pseudoMajorana1}, which could be understood  to be a Majorana fermion by 
chirality changing 
charge conjugation $\tilde{C}$ (pseudo-C)
 and chirality changing "parity" $\tilde{P}$, is shown not to be a 
Majorana fermion by the theorem on the absence of Majorana-Weyl 
fermions in d=4 \cite{Fujikawa}. The fermion number violating chiral fermion does not necessarily imply a Majorana fermion.

\section{Type I seesaw model} 

The Type I seesaw model
\cite{Seesaw} is defined by the neutrino sector  (in what follows, we  use the  conventions
in \cite{Bjorken} by setting $\gamma_{4}\rightarrow \gamma^{0}$ and $C\rightarrow i\gamma^{2}\gamma^{0}$)
\begin{eqnarray}\label{Seesaw}
{\cal L}_{\nu}&=&
\overline{\nu_{L}}(x)i\delslash\nu_{L}(x)
+\overline{\nu_{R}}(x)i\delslash\nu_{R}(x)\nonumber\\
&&- \{ \overline{\nu_{L}}m_{D}\nu_{R}(x)
+\frac{1}{v}\overline{\nu_{L}}m_{D}\nu_{R}(x)\phi(x)
+(1/2)\nu^{T}_{R}(x)Cm_{R}^{\dagger}\nu_{R}(x)+h.c.\}\nonumber\\
&&+\frac{g}{\sqrt{2}}\overline{l_{L}}^{k}(x)\gamma^{\mu}W_{\mu}^{-}(x)(\frac{1-\gamma_{5}}{2})U_{D}^{kl}{\nu_{L}^{l}}(x)
+ h.c.
\end{eqnarray}
where the lepton number violating dimension-3  right-handed mass term with $3\times 3$ $m_{R}$ is added to the model of Dirac neutrinos in an extension of the SM. This is consistent with the gauge symmetry of
the SM and gives very small masses of the order of $m_{D}^{2}/|m_{R}|$ to the measured
neutrinos by assuming that $\nu_{R}$ are gauge singlets
with large mass $m_{R}$. The Dirac-type mass
$m_{D}$ may be chosen to be a real diagonalized $3\times 3$ matrix. The
anti-symmetry of the charge conjugation matrix 
$C$ 
and Fermi statistics imply that $m_{R}$ is a complex and
symmetric $3\times 3$ matrix; the mass term is CP invariant for real $m_{R}$.
Eq.\eqref{Seesaw} is written in the  unitary gauge
with $v$ standing for the vacuum value of the Higgs
particle. We assume \eqref{Seesaw}, which is extremely left-right asymmetric, as our starting point without asking its derivation.
 
In the present analysis we adopt the chirality preserving C and P
for chiral fermions defined in \eqref{C and P}.
The CP and  parity violations in weak interactions are well described using C, P and CP thus defined.
We start with the mass term written as (by ignoring the
Higgs $\phi(x)$ term and weak interactions for a moment)
\begin{eqnarray}\label{mass term}
(-2){\cal L}_{mass}=
\left(\begin{array}{cc}
            \overline{\nu_{R}}&\overline{\nu_{R}^{C}}
            \end{array}\right)
\left(\begin{array}{cc}
            m_{R}& m_{D}\\
            m_{D}^{T}&0
            \end{array}\right)
            \left(\begin{array}{c}
            \nu_{L}^{C}\\
            \nu_{L}
            \end{array}\right) +h.c.,
\end{eqnarray}
where 
$\nu_{L}^{C} = C\overline{\nu_{R}}^T$ and $\nu_{R}^{C} =C\overline{\nu_{L}}^T$.
Since the above mass matrix is complex and symmetric, one can
diagonalize it precisely
by a $6 \times 6$ unitary $U$ (Autonne--Takagi
factorization\cite{Autonne--Takagi}) as
\begin{eqnarray}\label{orthogonal}
            U^{T}
            \left(\begin{array}{cc}
            m_{R}& m_{D}\\
            m_{D}^{T}& 0
            \end{array}\right)
            U
            =\left(\begin{array}{cc}
            M_{1}&0\\
            0&-M_{2}
            \end{array}\right)    ,        
\end{eqnarray}
where $M_{1}$ and $M_{2}$ are $3\times 3$ real diagonal
matrices. For an explicitly solvable case of single flavor, one can make $m_{R}$ real and 
$M_{1,2}=\sqrt{(m_{R}/2)^{2}+m_{D}^{2}} \pm m_{R}/2$. 

We can thus rewrite the seesaw Lagrangian  using 
\begin{eqnarray} \label{variable-change}          
            &&\left(\begin{array}{c}
            \nu_{L}^{C}\\
            \nu_{L}
            \end{array}\right)
            = U \left(\begin{array}{c}
            \tilde{\nu}_{L}^{C}\\
            \tilde{\nu}_{L}
            \end{array}\right)
           ,\ \ \ \ 
            \left(\begin{array}{c}
            \nu_{R}\\
            \nu_{R}^{C}
            \end{array}\right)
            = U^{\star} 
            \left(\begin{array}{c}
            \tilde{\nu}_{R}\\
            \tilde{\nu}_{R}^{C}
            \end{array}\right),         
\end{eqnarray}
which keep the kinetic terms in \eqref{Seesaw} invariant
and thus anti-commutation relations intact. Using the mass
eigenvalues $M_{1}$ and $M_{2}$, we have (by ignoring weak interactions)
\begin{eqnarray}\label{exact-solution}
{\cal L_{\nu}}
&=&(1/2)\{\overline{\psi_{+}}[i\delslash -M_{1}]\psi_{+}
+\overline{\psi_{-}}[i\delslash - M_{2}]\psi_{-}\} ,
\end{eqnarray}
 with (by suppressing the tilde symbols of $\tilde{\nu}$)
\begin{eqnarray}\label{pseudoMajorana}
\psi_{+}(x)= \nu_{R}+ C\overline{\nu_{R}}^{T},\ \
\psi_{-}(x)=\nu_{L}- C\overline{\nu_{L}}^{T}.
\end{eqnarray}
Note that the lepton number is not defined for $\psi_{\pm}$.
The Lagrangian \eqref{exact-solution} with real mass
parameters is CP invariant defined in \eqref{CP}, and the
possible CP breaking contained in the matrix $U$ is
transferred to the PMNS mixing matrix $U_{D}$ in \eqref{Seesaw}.  

The fields $\psi_{+}(x)$ and
$\psi_{-}(x)$ are  transformed by the charge conjugation \eqref{C and P}
\begin{eqnarray}
\psi_{+}(x)^{\hat{C}}&=& \nu_{L}+ C\overline{\nu_{L}}^{T}\neq 
\psi_{+}(x),\nonumber\\
\psi_{-}(x)^{\hat{C}}&=&-\nu_{R}+ C\overline{\nu_{R}}^{T}\neq 
-\psi_{-}(x),
\end{eqnarray}
while they are 
formally invariant under the commonly used {\em
chirality changing pseudo} $\tilde{C}$ \cite{Bilenky} and chirality changing 
$\tilde{P}$ (to maintain $\tilde{C}\tilde{P}=\hat{C}\hat{P}$),
\begin{eqnarray}\label{pseudo-C and P}
\tilde{C}&:& \ \nu_{L}(x)\rightarrow \nu^{\tilde{C}}_{L}(x)=
C\overline{\nu_{L}}^{T}(x), 
\ \ \nu_{R}(x)\rightarrow \nu^{\tilde{C}}_{R}(x)= 
C\overline{\nu_{R}}^{T}(x),\nonumber\\
\tilde{P}&:& \ \nu_{L}(x)\rightarrow\nu^{\tilde{P}}_{L}(x)= 
i\gamma^{0}\nu_{L}(t,-\vec{x}), 
\ \ \nu_{R}(x)\rightarrow \nu^{\tilde{P}}_{R}(x) =i\gamma^{0}\nu_{R}(t,-\vec{x}).
\end{eqnarray}
For example, $\psi_{+}(x)$ is invariant under the pseudo-C 
\begin{eqnarray}
\psi_{+}(x)^{\tilde{C}}=  C\overline{\nu_{R}}^{T} +\nu_{R}=\psi_{+}(x),
\end{eqnarray}
 but pseudo-C is not defined as an operator $\nu_{R}^{\tilde{C}}=((\frac{1+\gamma_{5}}{2})\nu_{R})^{\tilde{C}}= (\frac{1+\gamma_{5}}{2})\nu^{\tilde{C}}_{R}=(\frac{1+\gamma_{5}}{2})C\overline{\nu_{R}}^{T}=0$. This is a
consequence of the theorem on the absence of Majorana-Weyl
fermions in d=4 \cite{Fujikawa}. Also, a weird relation
(i.e., a chiral fermion is obtained by a chiral projection
of a Majorana fermion instead of a Dirac fermion)
\begin{eqnarray}
(\frac{1-\gamma_{5}}{2})\psi_{-}(x)=\nu_{L}(x)
\end{eqnarray}
does not lead to the neutrinoless double  $\beta$ decay if one should identify $\psi_{-}$ with a Majorana neutrino by the chirality changing pseudo-C \cite{Fujikawa}.
One may also note the parity (mirror symmetry)  in \eqref{C and P};  $\psi_{+}(x)^{\hat{P}}\neq i\gamma^{0}\psi_{+}(t,-\vec{x})$ and $\psi_{-}(x)^{\hat{P}}\neq i\gamma^{0}\psi_{-}(t,-\vec{x})$.
Although CP is well defined;  $\psi_{\pm}^{CP}(x)=\pm i\gamma^{0} \psi_{\pm}(t,-\vec{x})$.

\section{Conceptual basis of leptogenesis}
The simplest picture of the leptogenesis \cite{Fukugita-Yanagida, Liu-Segre, Covi}   is to generate the lepton number asymmetry 
starting with very massive Majorana fermions in the 
seesaw model \eqref{Seesaw}.  The CP violation is provided by  
 the complex heavy right-handed mass terms before diagonalization. 
The Majorana fermions imply that they are well-defined under C and P, and thus under CP. 
However, the fermions appearing in \eqref{pseudoMajorana} are 
largely C and 
P violating although CP is well-defined for real $M_{\pm}$ (for Dirac neutrinos with $m_{R}=0$, both C and P are 
well-defined).  
 The heavy fermions $\psi_{+}$ are the lepton number violating linear combinations of chiral fermions $\nu_{R}$ and 
$C\overline{\nu_{R}}^{T}$, and they are not  
conventional Majorana fermions. 

As a way
to allow a neutrinoless double $\beta$ decay at low energies mediated
 by the fermions $\psi_{-}$, it was suggested in \cite{Fujikawa} to apply a generalized Pauli-Gursey canonical transformation  \cite{Fujikawa2019}
 (analogue of the Bogoliubov canonical transformation \cite{Bogoliubov}), which mixes a neutrino and an 
antineutrino,  to those fermions in 
\eqref{pseudoMajorana}, thereby rendering  $\psi_{-}$ to be the 
Majorana fermions with a  change of the definition of the 
vacuum.
We thus consider a  $6 \times 6$ generalized
Pauli-Gursey canonical transformation $O$ with an {\em
invertible} orthogonal matrix
\begin{eqnarray}\label{Pauli-Gursey}
O=\frac{1}{\sqrt{2}} \left(\begin{array}{cc}
            1&1\\
            -1&1
            \end{array}\right),
\end{eqnarray}
which preserves CP.  By
suppressing the tilde symbols of $\tilde{\nu}$ in \eqref{variable-change}, we have
\begin{eqnarray} \label{Pauli--Gursey0}                   
            &&\left(\begin{array}{c}
            \nu_{L}^{C}\\
            \nu_{L}
            \end{array}\right)
            = O \left(\begin{array}{c}
            N_{L}^{C}\\
            N_{L}
\end{array}\right) =\frac{1}{\sqrt{2}}\left(\begin{array}{c}
N_{L}^{C}+N_{L}\\
            - N_{L}^{C} +N_{L}
            \end{array}\right) 
           ,\nonumber\\ 
           &&  \left(\begin{array}{c}
            \nu_{R}\\
            \nu_{R}^{C}
            \end{array}\right)
            = O 
            \left(\begin{array}{c}
            N_{R}\\
            N_{R}^{C}
\end{array}\right)=\frac{1}{\sqrt{2}}\left(\begin{array}{c}
N_{R}^{C}+N_{R}\\
            - N_{R}^{C} +N_{R}
            \end{array}\right).
\end{eqnarray}
The seesaw Lagrangian \eqref{exact-solution} is then
transformed to \cite{Fujikawa-Tureanu}
\begin{eqnarray}\label{seesaw Lagrangian3}
{\cal L_{\nu}}
&=&(1/2)\{\overline{\Psi_{+}}[i\delslash -M_{1}]\Psi_{+}
+\overline{\Psi_{-}}[i\delslash - M_{2}]\Psi_{-}\},
\end{eqnarray}
where 
\begin{eqnarray}\label{Pauli-Gursey2}
\psi_{+}&=&\nu_{R}+C\overline{\nu_{R}}^{T}\rightarrow
\frac{1}{\sqrt{2}} (N +N^{C}) \equiv \Psi_{+},\nonumber\\
\psi_{-}&=&\nu_{L}-C\overline{\nu_{L}}^{T}\rightarrow
\frac{1}{\sqrt{2}}(N -N^{C}) \equiv \Psi_{-}.
\end{eqnarray}
The transformed Type I seesaw Lagrangian \eqref{seesaw
Lagrangian3}, where no $\gamma_{5}$ appears,
is invariant under the standard C, P and CP defined
by 
\begin{eqnarray}\label{standard C and P}
&&\hat{C}:\ \ N(x)\leftrightarrow
N^{C}(x)=C\overline{N}^{T}(x),\nonumber\\
&&\hat{P}:\ \ N(t,\vec{x})\rightarrow i\gamma^{0}N(t,-\vec{x}),\ \
N^{C}(t,\vec{x})\rightarrow
i\gamma^{0}N^{C}(t,-\vec{x}),\nonumber\\
&&\hat{C}\hat{P}:\ \ N(x)\rightarrow i\gamma^{0}N^{C}(t,-\vec{x}),\ \
N^{C}(x)\rightarrow i\gamma^{0}N(t,-\vec{x}),
\end{eqnarray}
namely, $N(t,\vec{x})$ are {\em Dirac-type} variables and 
$\Psi_{\pm}$ are conventional Majorana fermions;  $N(t,\vec{x})$ satisfy the Dirac equations with lepton number violating terms \cite{Fujikawa2019}. 

The fermions in 
\eqref{pseudoMajorana} are thus converted to the conventional Majorana fermions $\Psi_{\pm}$
in \eqref{Pauli-Gursey2} by changing 
the definition of the neutrino vacuum  in d=4, thus ensuring the existence of the neutrinoless double $\beta$ decay \cite{Fujikawa}.  For $m_{R}=0$ (and then $M_{1}=M_{2}$), we have well-defined Dirac neutrinos in \eqref{seesaw Lagrangian3} consistent with \eqref{Seesaw}.  This canonical transformation is analogous to the Bogoliubov transformation in BCS theory.

\section{Leptogenesis and Higgs coupling}

It is sketched that the Majorana fermions $\Psi{\pm}$ thus defined describe naturally the leptogenesis. The Lagrangian in
\eqref{Seesaw} is rewritten in terms of Majorana neutrinos
\begin{eqnarray}\label{seesaw Lagrangian5}
{\cal L}_{\nu}&=&(1/2)\{\overline{\Psi_{+}}[i\delslash
-M_{1}]\Psi_{+} +\overline{\Psi_{-}}[i\delslash -
M_{2}]\Psi_{-}\}\nonumber\\
&&+\frac{g}{\sqrt{2}}\overline{l_{L}}^{k}(x)\gamma^{\mu}W_{\mu}^{-}(x)(\frac{1-\gamma_{5}}{2})[U^{k
l}{\Psi_{-}^{l}}(x)+\tilde{U}^{k l}{\Psi_{+}^{l}}(x)] + h.c.
\nonumber\\
&&-\phi(x) \{
(\Psi_{+}^{T}U_{11}^{T}+(\Psi_{-}^{C})^{T}U_{12}^{T})C(\frac{1-\gamma_{5}}{2})g_{h}[(U_{22}\Psi_{-}+U_{21}\Psi_{+})]
\nonumber\\
&&\hspace{0.5cm}+
(\Psi_{+}^{T}U_{11}^{\dagger}+\Psi_{-}^{T}U_{12}^{\dagger})C(\frac{1+\gamma_{5}}{2})g_{h}[(U^{\star}_{22}\Psi^{C}_{-}+U^{\star}_{21}\Psi_{+})]\}
\end{eqnarray}
where $g_{h}=m_{D}/v$ is the real diagonalized Higgs
coupling matrix, and $\Psi_{+}$ and $\Psi_{-}$ are the
Majorana neutrinos defined in \eqref{Pauli-Gursey2};
$\Psi_{+}^{C}=\Psi_{+}$ and $\Psi_{-}^{C}=-\Psi_{-}$.
The last two Higgs couplings in \eqref{seesaw Lagrangian5}
come from the Higgs couplings in \eqref{Seesaw} defined by the unitary 
gauge
${\cal L}_{H} =
-[\frac{1}{v}\overline{\nu_{R}}m_{D}\nu_{L}(x)\phi(x) +
h.c.]$: They are obtained by the diagonalization of neutrino masses in
\eqref{variable-change} combined with the Bogoliubov-type
transformation in \eqref{Pauli--Gursey0}, namely,
\begin{eqnarray}
\nu_{R}=U^{\star}_{11}\tilde{\nu}_{R} +
U^{\star}_{12}\tilde{\nu}_{R}^{C} & \rightarrow &
\frac{(1+\gamma_{5})}{2}[U^{\star}_{11}{\Psi_{+}}+
U^{\star}_{12}{\Psi_{-}}] \nonumber\\
\nu_{L}=U_{21}\tilde{\nu}_{L}^{C} + U_{22}\tilde{\nu}_{L} &
\rightarrow & \frac{(1-\gamma_{5})}{2}[U_{21}{\Psi_{+}} +
U_{22}{\Psi_{-}}].
\end{eqnarray}
The Lagrangian \eqref{seesaw Lagrangian5}
is exact as arising from a series of
well-defined transformations.
The CP breaking in weak interactions, which is determined
by $U^{kl}=U_{D}^{km}(U_{22})^{m l}$ (and $\tilde{U}^{kl} = U^{km}_{D} (U_{21})^{ml}$), and the CP breaking in
the effective Higgs couplings, which is determined by the
$6\times 6$ matrix $U$ in \eqref{variable-change} (mainly, 
$3\times 3$ submatrices
$U_{11}$ and $U_{22}$), are in principle independent.

In the Higgs couplings in \eqref{seesaw Lagrangian5}, both
$\hat{C}$ and $\hat{P}$ in \eqref{C and P} are broken but CP in \eqref{CP} is
preserved for {\em real} $U$. Moreover, the lepton number is not defined in the Higgs couplings in
\eqref{seesaw Lagrangian5} if one recalls that the phases
of the Majorana fermions are fixed by the (necessary
conditions of Majorana) $C\overline{\Psi_{+}}^{T}=\Psi_{+}$ and
$C\overline{\Psi_{-}}^{T}=-\Psi_{-}$. This absence of the
 lepton number persists even for the {\em complex} $U$, 
which breaks CP.

  Based on the conventional Majorana  propagator and the
conventional Higgs propagator in the unitary gauge with complex $U$, one can
evaluate one-loop diagrams using {\em only} the Higgs
couplings in \eqref{seesaw Lagrangian5} for the decay of a
heavy Majorana neutrino into a light Majorana neutrino plus
a Higgs particle. Including the interferences of the tree
diagrams and one-loop diagrams \cite{Fukugita-Yanagida,Liu-Segre, Covi}, we have a probability amplitude 
\begin{eqnarray}\label{physical matrix}
\langle \phi, (\frac{1\pm
\gamma_{5}}{2})\Psi_{-}|S|\Psi_{+}\rangle
\end{eqnarray} 
for $\Psi_{+} \rightarrow \Psi_{-} + \phi$,
producing  the chirality asymmetry among 
$(\frac{1\pm\gamma_{5}}{2})\Psi_{-}$ in the final states 
due to the CP breaking in the Higgs couplings, but no lepton
number asymmetry  in the unitary gauge. To incorporate the lepton number violation, one needs to convert the chirality 
asymmetry in \eqref{physical matrix} to the lepton  number asymmetry by the subsequent 
weak interactions \cite{Liu-Segre}.
   At far off-shell (at the
regions where the small masses are not important), the
chirality asymmetry of the Majorana neutrinos induce
correlations
\begin{eqnarray}
(\frac{1\pm\gamma_{5}}{2})\Psi_{-} \rightarrow
C(\frac{1\mp\gamma_{5}}{2})l^{\mp} +W^{\pm}
\end{eqnarray}
if one recalls the weak interaction vertices in
\eqref{seesaw Lagrangian5}.
In the present unitary gauge, it is essential to go through
the weak interactions to realize the observable lepton
number asymmetry \cite{Liu-Segre}.

We thus realize a scenario that the decay of heavy Majorana 
neutrinos $\Psi_{+}$ can produce the lepton number asymmetry \cite{Fukugita-Yanagida}.
\\

\noindent {\bf \large Landau gauge formulation}\\
\\
The leptogenesis is usually analyzed without any specific electroweak gauge fixing, particularly at finite temperatures
\cite{Fukugita-Yanagida, Covi}. Such a case may be described by  
the Landau gauge by adding the gauge fixing Lagrangian
\begin{eqnarray}
{\cal
L}_{\rm Landau}=-\frac{\alpha}{2}(\partial_{\mu}A^{\mu})^{2}-\frac{\beta}{2}(\partial_{\mu}Z^{\mu})^{2}-\gamma|\partial_{\mu}W^{\mu}|^{2}
+\ {\rm Faddeev-Popov\ terms},
\end{eqnarray}
with  real constants $\alpha, \beta\  {\rm and}\  \gamma$.  Note that the gauge 
fixing is required in any calculations in quantum theory. 
The
electromagnetic charge conservation is
maintained after this gauge fixing. The Higgs doublet after
the spontaneous symmetry breaking
 is parameterized by
\begin{eqnarray}\label{Higgs}
H(x)\rightarrow \left(\begin{array}{c}
            H^{+}(x)\\
            H^{0}(x) +\frac{1}{\sqrt{2}}v
            \end{array}\right) 
             =\frac{1}{\sqrt{2}}\left(\begin{array}{c}
            \varphi^{1}(x)+i\varphi^{2}(x)\\
            \phi(x)+i\varphi^{0}(x) +v
            \end{array}\right).
\end{eqnarray}
The unitary gauge \eqref{seesaw Lagrangian5} is defined by
setting $\varphi^{1}(x)=\varphi^{2}(x)=\varphi^{0}(x)=0$.
In contrast, the last two terms in
\eqref{seesaw Lagrangian5} are replaced  in the Landau gauge by
\begin{eqnarray}\label{seesaw Lagrangian7}
&&{\cal L}_{H}=-\sqrt{2}\{
(\Psi_{+}^{T}U_{11}^{T}+(\Psi_{-}^{C})^{T}U_{12}^{T})C(\frac{1-\gamma_{5}}{2})g_{h}[H^{0}(U_{22}\Psi_{-}+U_{21}\Psi_{+})-H^{+}
l] \nonumber\\
&&\hspace{0.3cm}+
(\Psi_{+}^{T}U_{11}^{\dagger}+\Psi_{-}^{T}U_{12}^{\dagger})C(\frac{1+\gamma_{5}}{2})g_{h}[(H^{0})^{\dagger}(U^{\star}_{22}\Psi^{C}_{-}+U^{\star}_{21}\Psi_{+})-H^{-}
l^{C}]\}.
\end{eqnarray}
The lepton number in this Higgs coupling is broken (by the
terms containing $H^{\pm}$) and C is also violated, but CP
is preserved for a {\em real} $6\times 6$ matrix $U$ in
\eqref{variable-change}.  At high temperatures T, one would set effectively $v\rightarrow 0$
in \eqref{seesaw Lagrangian5} and \eqref{seesaw Lagrangian7}.

As for the lepton number violation, the asymmetry of charged leptons $l$ and 
$l^{C}$ is always 
accompanied by the same asymmetry of the
charged Higgs scalars $H^{\pm}$ because of the
electromagnetic charge conservation in \eqref{seesaw Lagrangian7}. The later fate of $H^{\pm}$ thus becomes 
important. For the
{\em complex} $6\times 6$ U, CP is broken and two of the
3 necessary conditions of Sakharov are satisfied (except
for the thermal non-equilibrium, which is not considered in
our analysis). In our discussions so far,
the Higgs terms generating lepton number asymmetry in the leptogenesis are not electroweak gauge
independent. The
estimate of the on-shell S-matrix elements
\begin{eqnarray}
\langle H^{\pm}, (\frac{1\mp \gamma_{5}}{2})
l^{\mp}|S|\Psi_{+}\rangle
\end{eqnarray}
containing charged Higgs scalars $H^{\pm}$, which are
excluded as unphysical in the conventional treatment
at $T=0$ by the BRST cohomology, for example,  would  require  due care \cite{Fukugita-Yanagida, Covi}. The physical amplitude
\eqref{physical matrix}, which by itself does not induce explicit lepton
number violation, is still defined \cite{Liu-Segre}, although a careful treatment is required for the calculations at  finite temperatures.

\section{Conclusion}
One cannot assign an  ordinary physical meaning 
 to the  "neutrinos" \eqref{pseudoMajorana}
 in the Type I seesaw model in the framework of  conventional field 
theory in d=4 \cite{Fujikawa}, unlike the natural formulation of the SM with $\nu_{R}=0$ or the Dirac neutrinos with $m_{R}=0$.
To describe the neutrinoless double 
$\beta$ decay and the natural leptogenesis in the Type I seesaw 
model, it is essential to define Majorana fermions by 
introducing a new vacuum using an analogue of the  Bogoliubov canonical transformation.

\section*{Acknowledgements}

I thank J. Arafune, K. Hikasa and S. Iso for valuable comments.


\begin{thebibliography}{99}
 \bibitem{Majorana}
E. Majorana, "Teoria symmetrica dell'elettrone e del positrone'', 
Nuovo Cim. \textbf{14} (1937) 171 (An English translation by L. Maiani,
 ``A symmetric theory of electrons and positrons'', Soryusiron Kenkyu 
{\bf 63} (1981) 149. ).
\bibitem{Fujikawa}
K. Fujikawa, ``Absence of Majorana-Weyl fermions in d=4 and the 
theory of Majorana fermions'', Phys. Rev. D{\bf 113}, 013001 (2026).  

\bibitem{Schechter}
J. Schechter and J.W.F. Valle,
``Neutrino Masses in SU(2) x U(1) Theories,''
Phys. Rev. D \textbf{22} (1980) 2227.

\bibitem{CPT}
G. Luders, "On the equivalence of invariance under time reversal and 
under particle-antiparticle conjugation for relativistic field theories", Dan. Mat. Fys. Medd. {\bf 28} (1954) 1–17.\\
W. Pauli,  "Exclusion principle, Lorentz group and reflection of 
space-time and charge", Niels Bohr and the Development 
of Physics, W. Pauli (Ed.),  (London, Pergamon Press, 1955) pp. 30–51. 

\bibitem{Pauli}
W. Pauli,
``On the conservation of the Lepton charge,''
Nuovo Cim. \textbf{6} (1957) 204.\\
F. Gursey,
``Relation of charge independence and baryon conservation to 
Pauli\textquoteright{}s transformation,''
Nuovo Cim. \textbf{7} (1958) 411.

\bibitem{Seesaw} 
P. Minkowski, ``$\mu\rightarrow e\gamma$ at a rate of one out of 
$10^{9}$ muon decays?'', Phys. Lett. B{\bf 67} (1977) 421.\\
T. Yanagida, in Proceedings of Workshop on Unified Theory and Baryon
 Number
in the Universe, ed. by O. Sawada and A. Sugamoto 
(KEK report 79-18,
1979), p. 95.\\
M. Gell-Mann, P. Ramond and R. Slansky, in Supergravity, ed. by 
P. van
Nieuwenhuizen and D.Z. Freedman (North-Holland, Amsterdam, 
1979), p. 315.\\
R. N. Mohapatra and G. Senjanovic, ``Neutrino Masses and Mixings in 
Gauge Models With Spontaneous Parity Violation ``, Phys. Rev. Lett. 
{\bf 44} (1980) 912.

\bibitem{Bjorken}
J.D. Bjorken and S. D. Drell, {\it Relativistic Quantum Fields} (McGraw Hill, New York, 1965).


\bibitem{Autonne--Takagi}
L. Autonne, ``Sur les matrices hypohermitiennes et sur les matrices 
unitaires'', Ann. Univ. Lyon {\bf 38} (1915) 1.\\
T. Takagi, ``On an algebraic problem related to an analytic theorem of 
Caratheodory and Fejer and on an allied theorem of Landau'', 
Jpn. J. Math. {\bf 1} (1925) 83.

\bibitem{Bilenky}
S. M. Bilenky, J. Ho\v sek and S. T. Petcov,
``On Oscillations of Neutrinos with Dirac and Majorana Masses,''
Phys. Lett. B \textbf{94} (1980) 495.\\
M. Doi, T. Kotani, H. Nishiura, K. Okuda and E. Takasugi,
``CP Violation in Majorana Neutrinos,''
Phys. Lett. B \textbf{102} (1981) 323.

\bibitem{Fukugita-Yanagida}
M. Fukugita and T. Yanagida, ``Baryogenesis Without
Grand Unification," Phys. Lett., B{\bf 174} (1986) 45,
\bibitem{Liu-Segre}
J. Liu and G. Segre, ``Reexamination of generation of baryon and 
lepton number asymmetries in the early Universe by heavy particle 
decay'', Phys. Rev. D{\bf 48} (1993) 4609.

 \bibitem{Covi}
L. Covi, E. Roulet and F. Vissani,  ``CP violating decays in leptogenesis 
scenarios'', Phys.Lett. B{\bf 384} (1996) 169.


\bibitem{Fujikawa2019}
K. Fujikawa,
``Generalized Pauli\textendash{}Gursey transformation and Majorana 
neutrinos,''
Phys. Lett. B \textbf{789} (2019) 76. See also \cite{Pauli}.

\bibitem{Bogoliubov}
N. N. Bogoliubov, "A new method in the theory of superconductivity. I", Soviet Physics JETP {\bf 34} (1958) 41.

\bibitem{Fujikawa-Tureanu}   
K. Fujikawa and A. Tureanu, 
``Two classes of Majorana neutrinos in the seesaw model,''
Phys. Lett. B{\bf 858} (2024) 139064; ``Majorana neutrino as Bogoliubov quasiparticle,''
Phys. Lett. B{\bf 774} (2017) 273.

\end{thebibliography}
\end{document}